\documentclass[draftclsnofoot, letterpaper, onecolumn,12pt]{IEEEtran}
\usepackage{amsfonts}
\usepackage{amsmath,cases,amssymb}
\usepackage[dvips]{graphicx}
\usepackage{cite,color}
\usepackage{color}
\usepackage{subcaption}
\usepackage{epsfig}
\usepackage{epstopdf}
\usepackage{url}
\usepackage{algorithm}
\usepackage{algorithmicx, algpseudocode}
\usepackage{color}
\usepackage{multirow}
\usepackage{mathtools}
\usepackage{balance}
\newcommand{\rank}{{\sf rank}}

\newcommand{\clL}{{\cal L}}

\newcommand{\ds}{\displaystyle}

\newcommand{\clN}{{\cal N}}
\newcommand{\clG}{{\cal G}}

\newcommand{\Tr}{\mbox{Trace}}

\newcommand{\bgeqn}{\begin{equation}}
\newcommand{\edeqn}{\end{equation}}

\newcommand{\beqa}{\begin{eqnarray}}
\newcommand{\eeqa}{\end{eqnarray}}
\newcommand{\beqas}{\begin{eqnarray*}}
\newcommand{\eeqas}{\end{eqnarray*}}

\newcommand{\clT}{{\cal T}}
\newcommand{\clP}{{\cal P}}
\newcommand{\clV}{{\cal V}}

\newcommand{\clR}{{\cal R}}
\begin{document}
\title{Bang-Bang Charging of Electrical Vehicles by Smart Grid Technology}
\author{Y. Shi, H. D. Tuan, T. Q. Duong, H. V. Poor and  A. V. Savkin, 
\thanks{Ye Shi and Hoang D. Tuan are with the School of Electrical and Data Engineering, University of Technology Sydney, Broadway, NSW 2007, Australia (email: Ye.Shi@student.uts.edu.au, Tuan.Hoang@uts.edu.au)}
\thanks{Trung Q. Duong is with Queen's University Belfast, Belfast BT7 1NN, UK  (email: trung.q.duong@qub.ac.uk)}
\thanks{H. Vincent Poor is with the Department of Electrical Engineering, Princeton University, Princeton, NJ 08544, USA (e-mail: poor@princeton.edu)}
\thanks{Andrey V. Savkin with the School of Electrical Engineering and Telecommunications, The University of New South Wales
Sydney, NSW 2052, Australia (email: a.savkin@unsw.edu.au)   }
}
\date{}

\maketitle
\begin{abstract} The success of the transportation electricification in this century particularly requires
the penentration of the internet of plug-in electric vehicles (PEVs) into the smart power grid. Beside the function
of serving the  traditional residential power demand, next-generation  power grids  also aim to support
the internet of PEVs at the same time. The distinct difference between the traditional power demand and PEVs' power demand
is that while the statistics of the former is rich enough for treating it as inelastic/known before hand,
the latter is unknown until random PEVs' arrivals. Massive penentration of PEVs certainly causes the
grid unpredictable fluctuation. The present paper considers the joint PEVs charging coordination and grid power generation
to minimizing both of the negative impact of PEVs' integration and  the cost of power generation
while meeting the grid operating constraints and all parties' demand. The bang-bang
PEVs charging strategy is adopted to exploit its simple implementation. By using  a recently developed
model predictive control (MPC) model for this problem, the online compuation is based on a predictive
mixed integer nonlinear programming (MINP). A new solution computation for this optimization problem is developed.
Its capacity of achieving the  globally optimal solution is shown by numerical comparison between its performance
and that by an off-line optimal solution.
\end{abstract}
\section{Introduction}
Due to the increasing awareness of energy consumption and environment pollution from traditional fossil fuel, as well as the development of battery and charging technology, there will be a significant growing number of plug-in electric
vehicles (PEVs) within the next few years\cite{RI11}. The PEVs will play an important role in the future smart grid because of the benefits such as, lower operation cost, less air pollution emissions and better utilization of renewable energy\cite{JTG13}. However, as most PEVs utilize grid power for charging, the growing penetration of PEV could pose potential threats to the existing smart grid. Unregulated charging of PEV may lead to serious overloading, additional power loss and unacceptable voltage violation in smart grid system\cite{SHMV11,HSX12}. Therefore, optimal scheduling for PEV charging aims at
minimizing the total cost of PEVs and smart grid, while satisfying the charging demand of PEV and operation constraints of smart grid is necessary to study.

Recently, various works have addressed the optimal scheduling problem for PEVs in smart grid\cite{SE11,FRR15,AFRR16,Xietal16}. A mixed integer linear programming (MILP) formulation was proposed in\cite{HWZ14} to optimize the daily cost on PEV charging involving with linear dc power flow constraint. The disadvantage of dc power flow is clearly especially in smart grid, due to the higher dc error and exclusion of  bus voltage and reactive power from the model. Reference \cite{FRR15} proposed a mixed integer nonlinear programming (MINLP) model for optimal scheduling of PEV in an unbalanced distributed system. The MINLP problem was then linearized to MILP problem by some linearlization techniques such as the first order Taylor expansion and piecewise linear approximation. As a result, its solution of MILP is not necessarily feasible to the original MINLP problem. In \cite{AFRR16}, a similar MILP model with \cite{FRR15} was proposed by adding a vehicle-to-grid (V2G) charging strategy, which allows PEV behaving as bidirectional power source to reduce the negative effect at peak time\cite{SE11}. However, \cite{CHD11} raised  concern about the cost and techniques for discharging of PEV. In addition, the above reference for PEV charging all applied static scheduling strategy, which assumed that information including PEV arrival time, departure time and initial SOC are given beforehand. However, it is not realistic to obtain all of those information in advance.

Model predictive control (MPC) approach for dynamic PEV scheduling has emerged as a promising solution to deal with the system dynamic and uncertainty. In \cite{TZ17}, a
MPC-based model was formulated to schedule PEV charging in a finite horizon, but the operation constraints of grid were not considered. Additionally, its assumption of PEV could be fully charged in only one time slot is unrealistic due to the physical limitation of PEV itself. A MILP model formulated over a rolling horizon window for energy storage control was
proposed in \cite{MSE14}, while the voltage balance was ignored. \cite{RSME16} presented a MILP-based MPC for integrated PEVs scheduling in microgrid. Three-types of PEV charging scenarios are provided, including bidirectional, unidirectional and one block charging. However, the stochastic optimization in \cite{RSME16} suffers from large computational cost.

In this paper, both the dynamic and static scheduling for PEV charging are studied. A novel MPC-based two-stage computational solution is proposed to iteratively solve the dynamic scheduling problem. The static scheduling scenario serves as a counterpart to investigate the optimality of the MPC-based solution of dynamic scenario. Extensive simulation results based on real electricity price and residential demand shows that the proposed method is effective and practical.

The rest of the paper is structured as follows. Section
II is devoted to the problem statement of the PEV charging scheduling problem, which is formulated as a MINLP model in section III. A dynamic computational
solution for dynamic scheduling scenario, using the proposed MPC-based two-stage approach is
developed in Section IV. A static computational solution for static scheduling scenario
is considered in Section V. Section VI provides the computational results and Section VI concludes the paper.

{\it Notation.} The notation used in this paper is standard. Particularly, $j$ is the imaginary unit,
$X^H$ is Hermitian transpose  of a vector/matrix $X$, $M\succeq 0$ for a Hermitian symmetric matrix $M$ means that
it is positive semi-definite, $\rank(M)$ and $\Tr(M)$ are the rank and trace of a matrix $M$, respectively. $\Re (\cdot)$ and $\Im (\cdot)$ are the real and imaginary parts of a complex quantity, and
 $a\leq b$ for two complex numbers $a$ and $b$ is componentwise understood, i.e. $\Re(a)\leq \Re(b)$ and $\Im(a)\leq \Im(b)$.
The cardinality of a set ${\cal C}$ is denoted by $|{\cal C}|$. $\lceil x \rceil$ is the smallest integer that is not less than x.

\section{Problem statement}

\begin{figure}[h]
\centering
\includegraphics[width=0.9 \columnwidth]{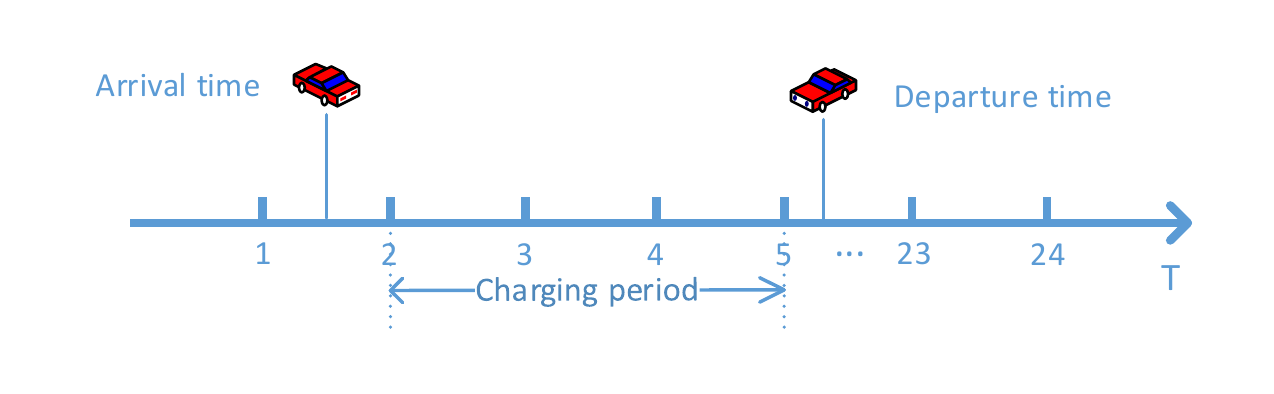}
\caption{Illustration of the charging period of a PEV}
\label{period}
\end{figure}

We consider a joint problem of PEV charging scheduling and power control in an residential grid, which aims at saving operation costs for both PEVs and power generation, with demands of PEVs charging and residential power grid satisfied. In the power grid, active and reactive power balance between supply and demand, physical limitations of grid including voltage and power bounds are taken into account. The serving time period of the grid is divided into $T$ time slots $\clT:=\{1, 2,\dots, T\}$. Each time slot has a time duration $\delta_t$, which usually varies from $30$ minutes to an hour. In addition, price-inelastic load varies from each time interval $[t,t+1]$ according to the residential specification. In this paper, the following are assumed:

\begin{itemize}
       \item The SOC of each PEV is known after plugged into the grid, and it must be fully charged by departure during the charing period $\clT$;
       \item The charging of PEV must be operated in a given period of time slots, Fig.\ref{period} illustrates the charging period of a PEV;
       \item The PEVs can communicate with the grid and  update the information of charging state, which can be controlled in each time interval during the charing period $\clT$.
\end{itemize}

Charging demands specified by charging tasks of PEVs and grid demands including active and reactive power balance between demand and supply, voltage and power limitations must be satisfied. In addition, we consider the objective as minimizing the total cost of PEVs charging and active power generation among the whole charging period $\clT$.
The charging state of a PEV can be represented by a binary variable $\tau\in\{0,1\}$. $\tau =1$ denotes the battery is charging at a fixed rate, while $\tau =0$ denotes the battery is not charging.

In this paper, both static and dynamic scheduling of PEV charging are studied. In the static charging scenario, all the information including PEV arrival time, departure time and initial SOC are given beforehand. The grid utility optimizes the variables of charging rate and active power by an algorithm and then all PEVs will follow the scheduling profile after plugged into the grid. However, it is not realistic to obtain all of those charging information in advance. Additionally, it is also not possible to have a stable long-term scheduling profile for the grid. Therefore, to tackle the PEV charging problem more realistic, it is significant to adopt the dynamic charging scenario, in which the utility calculates the optimal charging scheduling only for PEVs that connected with the grid at the current time slot. In the next time slot, the utility will update the charging information such as the set of PEVs connected in the grid, the SOC of PEVs and re-do the calculation. The above dynamic procedure is the basic idea of MPC approach. We will focus on the dynamic charging scenario in this work, and the static scenario will serve as a counterpart to investigate the optimality of our MPC-based dynamic charging method.

\section{MINLP Model formulation}

The joint problem of PEV charging scheduling and power control in an residential grid can be formulated as a MINLP problem. Consider a residential power grid with a set of buses ${\cal N} := \{1, 2,..., N\}$ connected
through a set of flow lines ${\cal L}\subseteq {\cal N}\times {\cal N}$, i.e. bus $k$ is connected to bus $m$ if and only if $(k,m)\in {\cal L}$. Accordingly, $\clN(k)$ is the set of other buses connected to bus $k$.
There is a subset ${\cal G}\subseteq {\cal N}$, whose elements are connected to distributed generators (DGs).
Any bus $k\in {\cal N}\setminus {\cal G}$ is thus not connected to DGs. Any bus $k\in {\cal G}$ also has a function
to serve PEVs and in what follow is also referred to  CS $k$. By defining $M=|{\cal G}|$, there are $M$ CSs in the grid. Denote by ${\cal H}_k$ the set of those PEVs that arrive at CS $k$. Accordingly, $k_n$ is the $n$-th PEV that arrives at CS $k$.
%
Following is a description of the MINLP model formulation, including constraints and objective function.

\subsection{Constraints of PEV Charging}
PEV $k_n$ arrives at $t_{a,k_n}\in \clT$ and needs to depart by $t_{k_n,d}\in\clT$. The constraint
\begin{equation}\label{time1}
t_{k_n,d}-t_{k_n,a}\leq T_{k_n},
\end{equation}
expresses the PEV $k_n$'s time demand.

Suppose that $C_{k_n}$ and $s_{k_n}^0$ are the battery capacity and initial SOC of PEV $k_n$.
It must be fully charged by the departure time $t_{k_n,d}$, i.e.
\begin{eqnarray}\label{PEV1b}
\ds\sum_{t'=t_{k_n,a}}^{t_{k_n,d}} u_h\bar{P}_{k_n}\tau_{k_n}(t') \geq  C_{k_n}(1 - s_{k_n}^{0}),
\end{eqnarray}
where $u_h$ is the charging efficiency of the battery, $\bar{P}_{k_{n}}(t')$ is fix charging rate of PEV $k_n\in {\cal H}_k$, $\tau_{k_n}(t')$ denotes the binary variables to represent the charging state of PEV $k_n\in {\cal H}_k$,
\begin{equation}\label{PEV1c}
\tau_{k_n}(t')= \left\{
\begin{array}{l}
1, \quad \text{If PEV $k_n$ is charging at time $t'$}; \\
0, \quad \text{If PEV $k_n$ is not charging at time $t'$}.
\end{array}
\right.
\end{equation}
For ease of presentation, if $t'\notin [t_{k_n,a}, t_{k_n,b}]$ we set $\tau_{k_n}(t')=0$.

\subsection{Constraints of Grid Limitations}
The next constraints relate to the acceptable range of generated power by the DGs:
\begin{eqnarray}
{\underline P}_{g_k} \leq P_{g_k}(t') \leq {\overline P}_{g_k}, \quad k\in {\cal G}, \label{PEV1d}
\end{eqnarray}
and
\begin{eqnarray}
{\underline Q}_{g_k} \leq Q_{g_k}(t') \leq {\overline Q}_{g_k}, \quad  k\in {\cal G}, \label{PEV1e}
\end{eqnarray}
where ${\underline P}_{g_k}$, ${\underline Q}_{g_k}$ and ${\overline P}_{g_k}$, ${\overline Q}_{g_k}$ are respectively the lower and upper limit of the real generated  and reactive generated powers, $P_{g_{k}}(t')$ and $Q_{g_{k}}(t')$ are the real
and reactive powers generated by DG $k$, respectively.

The constraints of voltage are
\begin{eqnarray}
{\underline V}_k \leq |V_k(t')| \leq {\overline V}_k, \quad  k\in {\cal N},\label{PEV1f}\\
|\mbox{arg}(V_k(t'))-\mbox{arg}(V_m(t'))| \leq \theta_{km}^{\max}, (k,m)\in\clL, t'\in\clT, \label{PEV1g}
\end{eqnarray}
where ${\underline V}_k$ and ${\overline V}_k$ are the lower limit and upper limit of the voltage amplitude,
while $\theta^{\max}_{k,m}$ are given to express the voltage phase balance.

\subsection{Constraints of PEV \& Power Balance}
Followed by \cite{Yeetal17}, for $k\in \clG$, the total supply and demand energy is balanced as
\begin{eqnarray}\label{PEV1h}
 V_k(t')(\sum_{m\in \clN(k)} y_{km} (V_k - V_m))^* =  (P_{g_{k}}(t') - P_{l_{k}}(t') \nonumber \\ -\sum_{n\in {\cal H}_k}\bar{P}_{k_n}\tau_{k_n}(t')) + j(Q_{g_{k}}(t') - Q_{l_{k}}(t')), k\in \clG,
\end{eqnarray}
for $k\in {\cal N}\setminus {\cal G}$,
\begin{eqnarray}\label{PEV1i}
V_k(t)(\sum_{m\in \clN(k)} y_{km} (V_k - V_m))^* = \nonumber \\  - P_{l_{k}}(t')  - j Q_{l_{k}}(t'),  k\in {\cal N}\setminus {\cal G},
\end{eqnarray}
where $y_{km}\in\mathbb{C}$ is the admittance of line $(k,m)$ and  $P_{l_{k}}(t')$ and $Q_{l_{k}}(t')$ are respectively
known real  and reactive price-inelastic demands
at bus $k$ to express the residential power demand

The problem of interest is to minimize both the energy cost to DGs and charging cost for PEVs. Thus, by defining
\[
V(t')=(V_1(t'),\dots, V_N(t')), \clV=\{V(t')\}_{t'\in\clT},
\]
\[
\begin{array}{c}
P_g(t')=(P_{g_1}(t'),\dots, P_{g_M}(t')),\\
Q_g(t')=(Q_{g_1}(t'),\dots, Q_{g_M}(t')),\\
R(t')=\{P_g(t'), Q_g(t')\},  \clR=\{R(t')\}_{t'\in\clT},
\end{array}
\]
and
\begin{eqnarray}
\clP^{PEV}=\{\tau^{PEV}(t')\}_{t'\in\clT}, \nonumber \\
\tau^{PEV}(t')=\{\tau_{k_n}(t')\}_{k_n\in{\cal H}_k, k=1,\dots, M}, \nonumber
\end{eqnarray}
the objective function is given by
\begin{equation}\label{objective1}
F(\clR,\clP^{PEV}) = \ds \sum_{t\in \clT} \sum_{k\in {\cal G}}f(P_{{g_k}}(t')) + \ds \sum_{t\in \clT} \sum_{k\in {\cal N}} \sum_{n\in {\cal H}_k} \beta_t\bar{P}_{{k_n}}\tau_{k_n}(t')
\end{equation}
where $f(P_{{g_k}}(t'))$ is the cost function of real power generation  by DGs, which is linear or quadratic in $P_{{g_k}}(t')$,
and $\beta_t$ is the known PEV charging price during the time interval $(t',t'+1]$.

The joint PEV charging scheduling and voltage control is mathematically formulated as
\begin{equation}\label{PEV1}
\ds\min_{\clV, \clR,\clP^{PEV}}\ F(\clR,\clP^{PEV})\quad
\mbox{s.t.}\quad  (\ref{PEV1b})-(\ref{PEV1i})
\end{equation}

The above problem (\ref{PEV1}) is MINLP, which is nonconvex and very computationally challenging because of the quadratic equality constraints
(\ref{PEV1h}) and (\ref{PEV1i}), nonlinear inequality constraints (\ref{PEV1f}) and (\ref{PEV1g}) and integer constraints (\ref{PEV1c}). Moreover, the arrival time $t_{k_n,a}$ of each individual PEV $k_n$, its charging
demand and its departure time $t_{k_n,d}$ are unknown. Generally, linearization techniques such as first order Taylor expansion and piece wise approximation are applied to handle the nonlinear constraints (\ref{PEV1h}), (\ref{PEV1i}), (\ref{PEV1f}) and (\ref{PEV1g}) in the above MINLP problem \cite{HWZ14,FRR15,AFRR16}. After the linearization, MILP model is obtained, which can be efficiently solved under the framework of commercial solver CPLEX. However, the found result are not necessarily feasible to the original problem (\ref{PEV1}) because the linearization techniques can not guarantee the consistence between the MINLP and MILP. In the next section, a novel MPC-based two-stage computational solution will be proposed to iteratively solve the MINLP problem (\ref{PEV1}).

\section{MPC-based two-stage computational solution for dynamic scheduling}

Considering $(R(t'), P^{PEV}(t'))$ and $V(t')$ as the plant state and control, respectively,
equations (\ref{PEV1d}), (\ref{PEV1e}), (\ref{PEV1h}) and (\ref{PEV1i}) provide
state behavioral  equations \cite{PW} with the end constraint (\ref{PEV1b}) together with  (\ref{PEV1b}), while equations (\ref{PEV1f})
and (\ref{PEV1g}) provide control constraints.  On the surface,
(\ref{PEV1}) appears to be a control problem over the finite horizon $[1,T]$.  However, all equations in (\ref{PEV1}) are unpredictable beforehand, preventing the application of conventional model predictive control \cite{Ca04,M16}. We now follow the
idea of \cite{Tuetal15} to address (\ref{PEV1}).

At each time $t$ denote by $C(t)$ the set of PEVs that need to be charged. For each $k_n\in C(t)$, let
$\clP_{k_n}(t)$ be its remaining demand for charging by the departure time $t_{k_n,d}$. Define
\begin{equation}\label{hori1}
\Psi(t)=\max_{k_n\in C(t)} t_{k_n,d}.
\end{equation}
At time $t$ we solve the following MINLP problem over
the prediction horizon $[t,\Psi(t)]$ but then take only
$V(t),  P_{k_n}(t),  R(t)$ for online updating solution of (\ref{PEV1}):
\allowdisplaybreaks[4]
\begin{subequations}\label{HOR1}
\begin{eqnarray}
\ds\min_{V(t'), R(t'), \tau_{k_n}(t'), t'\in [t,\Psi(t)], k_n\in C(t)} F_{[t,\Psi(t)]}&& \label{HOR1a}\\
\mbox{s.t.}\quad (\ref{PEV1d})-(\ref{PEV1g}), \quad (\ref{PEV1i}), \quad \mbox{for}\ t'\in [t,\Psi(t)],
 \label{HOR1g}\\
 V_k(t')(\sum_{m\in \clN(k)} y_{km} (V_k(t') - V_m(t')))^* = \nonumber \\
(P_{g_{k}}(t') - P_{l_{k}}(t') -\sum_{k_n\in C(t)}\bar{P}_{k_n}\tau_{k_n}(t'))
\nonumber\\ + j(Q_{g_{k}}(t') - Q_{l_{k}}(t')),  \quad  (t',k)\in [t,\Psi(t)]\times \clG,  \label{HOR1b}  \\
\ds\sum_{t'=t}^{t_{k_n,d}} u_h\bar{P}_{k_n}\tau_{k_n}(t') \geq \clP_{k_n}(t), t'\in [t,\Psi(t)],\label{HOR1h}\\
\tau_{k_n}(t')\in \{0,1\}, t'\in [t,\Psi(t)],\label{HOR1i}
\end{eqnarray}
\end{subequations}
with $F_{[t,\Psi(t)]}:=\ds \sum_{t'=t}^{\Psi(t)}
\sum_{k\in {\cal G}}f(P_{{g_k}}(t')) + \ds \sum_{t'=t}^{\Psi(t)} \sum_{k_n\in C(t)}\beta_t\bar{P}_{{k_n}}\tau_{k_n}(t') $.
One can notice that (\ref{HOR1}) includes only what is known at the present time $t$. Of course, (\ref{HOR1}) is still  a difficult MINLP and in the end we need only its solution
at $t$, so we propose the following approach in tackling its solution at $t$.

Define the Hermitian symmetric matrix  $W(t') = V(t')V^H(t') \in \mathbb{C}^{N\times N}$,
which must satisfy $ W(t')\succeq 0$ and $\rank(W(t'))=1$.
By replacing  $W_{km}(t')=V_k(t')V^*_m(t')$,  $(k,m)\in\clN\times\clN$ in,
we reformulate (\ref{HOR1}) to the following optimization problem in matrices $W(t')\in\mathbb{C}^{N\times N}$,
$t'\in [t,\Psi(t)]$:
\allowdisplaybreaks[4]
\begin{subequations}\label{rHOR1}
\begin{eqnarray}
\ds\min_{W(t'), R(t'), \tau_{k_n}(t'), t'\in [t,\Psi(t)], k_n\in C(t)} F_{[t,\Psi(t)]} \label{rHOR1a}\\
\mbox{s.t.}\quad (\ref{PEV1d})-(\ref{PEV1e}), \quad \mbox{for}\ t'\in [t,\Psi(t)],\label{rHOR1d}\\
\ds\sum_{m\in \clN(k)} (W_{kk}(t') - W_{km}(t'))y_{km}^*  = (P_{g_{k}}(t') - P_{l_{k}}(t')\nonumber \\
 -\sum_{k_n\in C(t)}\bar{P}_{k_n}\tau_{k_n}(t'))
 + j(Q_{g_{k}}(t') - Q_{l_{k}}(t')),  \quad  k\in \clG,  \label{rHOR1b}  \\
\sum_{m\in \clN(k)}(W_{kk}(t') - W_{km}(t'))y_{km}^* =   \nonumber\\
 - P_{l_{k}}(t')  - j Q_{l_{k}}(t'), k\notin  {\cal G}, \label{rHOR1c}\\
{\underline V}_k^2 \leq W_{kk}(t') \leq {\overline V}_k^2, \quad  k\in {\cal N}, \label{rHOR1f}\\
\Im(W_{km}(t'))\leq \Re(W_{km}(t'))\tan(\theta_{km}^{max}), (k,m)\in\clL,  \label{rHOR1g}\\
W(t')\succeq 0,  \label{rHOR1k}\\
\mbox{rank}(W(t'))=1,\label{rHOR1l}\\
\ds\sum_{t'=t}^{t_{k_n,d}} u_h\bar{P}_{k_n}\tau_{k_n}(t') \geq \clP_{k_n}(t), \label{rHOR1m}\\
\tau_{k_n}(t')\in \{0,1\}, \label{rHOR1n}
\end{eqnarray}
\end{subequations}

By now, the difficulty of (\ref{rHOR1}) is concentrated on the multiple nonconvex  matrix rank-one constraints (\ref{rHOR1k}) and the massive integer constraints (\ref{rHOR1m}) and (\ref{rHOR1n}). A two-stage nonsmooth algorithm is proposed to efficiently handle the two main difficulties step by step.

In the first stage, the matrix rank-one constraints (\ref{rHOR1k}) is relaxed, then we solve  the following problem to locate the solution of charging binary variable $\tau_{k_n}(t')$,
\begin{eqnarray}\label{rHOR2}
\ds\min_{W(t'), R(t'), \tau_{k_n}(t')} F_{[t,\Psi(t)]}\quad
 \mbox{s.t.}\ (\ref{rHOR1d})-(\ref{rHOR1k}), \quad (\ref{rHOR1m}),\quad (\ref{rHOR1n}).
\end{eqnarray}
Suppose that $(W^{(\kappa+1)}(t)$, $R^{(\kappa+1)}(t))$ and $\hat{\tau}_{k_n}(t')$ is the optimal solution of (\ref{rHOR2}). If
$\mbox{rank}(\hat{W}(t'))\equiv 1$, $t'\in [t,\Psi(t)]$, then $\hat{V}(t')$ such that $\hat{W}(t')=\hat{V}(t')\hat{V}^H(t')$
together with $\hat{R}(t')$ and  $\hat{P}_{k_n}(t')$ constitute
the optimal solution of the nonconvex optimization problem (\ref{HOR1}). Otherwise, we go to the next stage.

In the second stage by substituting $\hat{\tau}_{k_n}(t')$ into (\ref{rHOR1b}), we solve the following problem to obtain the solution of $W(t')$ and $R(t')$,
\begin{subequations}\label{rHOR3}
\begin{eqnarray}
\ds\min_{W(t'), R(t')}\ F(P_g(t'))):=\ds \sum_{t'=t}^{\Psi(t)}
\sum_{k\in {\cal G}}f(P_{{g_k}}(t')),\quad
 \mbox{s.t.}\ (\ref{rHOR1c})-(\ref{rHOR1l}), \label{rHOR3b}\\
\ds\sum_{m\in \clN(k)} (W_{kk}(t') - W_{km}(t'))y_{km}^*  = (P_{g_{k}}(t') - P_{l_{k}}(t')\nonumber \\
 -\sum_{k_n\in C(t)}\bar{P}_{k_n}\hat{\tau}_{k_n}(t'))
 + j(Q_{g_{k}}(t') - Q_{l_{k}}(t')),  \quad  k\in \clG,  \label{rHOR3c}
\end{eqnarray}
\end{subequations}
Following is the specific procedure to solve the two stages problem (\ref{rHOR2}) and (\ref{rHOR3}).

\subsection{The first stage: iterative procedure to solve (\ref{rHOR2})}
From (\ref{rHOR1m}) we can find an integer number $0<\bar{\tau}_{k_n}< t_{k_n,d}-t_{k_n,a}$ such that
\begin{equation}\label{time2}
\sum_{t=t_{k_n,a}}^{t_{k_n,b}}\tau_{k_n}(t)=\bar{\tau}_{k_n}:=\lceil\frac{C_{k_n}(1 - s_{k_n}^{0})}{u_h\bar{P}_{k_n}}\rceil,
\end{equation}
where $\lceil x\rceil$ is the smallest integer that is not less than $x$.
The discrete constraints (\ref{rHOR1m}) and (\ref{rHOR1n}) are equivalent to the following set of continuous constraints
\begin{eqnarray}
(\ref{time2}),\ \tau_{k_n}(t)\in [0,1], t\in [t_{k_n,a},t_{k_n,b}],\label{time3}\\
g(\tau)=\ds \sum_{t=t_{k_n,a}}^{t_{k_n,b}} \sum_{k_n\in C(t)}\geq \bar{\tau}:=
\sum_{k_n\in C(t)}\bar{\tau}_{k_n}\label{time4}
\end{eqnarray}
for any $L>1$ (generally $L$ is set as $1.5$). The difficulty is now concentrated on the reverse convex constraint (\ref{time4}).
We then address (\ref{rHOR2}) by
\begin{subequations}\label{rHOR4}
\begin{eqnarray}
\ds\min_{W(t'), R(t'), \tau_{k_n}(t')} F_{[t,\Psi(t)]} + \mu_1(\ds\frac{1}{g(\tau)}-\frac{1}{\bar{\tau}})\label{rHOR4a}\\
\mbox{s.t.}\quad\quad (\ref{rHOR1d})-(\ref{rHOR1k}), \quad (\ref{time3}),\quad (\ref{time4}), \label{rHOR4b}
\end{eqnarray}
\end{subequations}
where $\mu_1>0$ is a penalty parameter.

At each iteration $\kappa$, we solve
\begin{subequations}\label{rHOR5}
\begin{eqnarray}
\ds\min_{W(t'), R(t'), \tau_{k_n}(t')} F_{[t,\Psi(t)]} + \mu_1(\ds\frac{1}{g^{(\kappa)}(\tau)}-\frac{1}{\bar{\tau}})\label{rHOR5a}\\
\mbox{s.t.}\quad (\ref{rHOR4b}), L \tau_{k_n}(t)\geq (L-1)\tau^{(\kappa)}_{k_n}(t),\label{rHOR5b}
\end{eqnarray}
\end{subequations}
where $g^{(\kappa)}$ is the linearization of $g$ based on the first order Taylor expansion at $\tau^{(\kappa)}$:
\begin{equation}\label{gk}
g^{(\kappa)}(\tau)=
L\sum_{t=t_{k_n,a}}^{t_{k_n,b}}\sum_{k_n\in C(t)}(\tau^{(\kappa)}_{k_n}(t))^{L-1}\tau_{k_n}(t)
-(L-1)\sum_{t=t_{k_n,a}}^{t_{k_n,b}}\sum_{k_n\in C(t)}(\tau^{(\kappa)}_{k_n}(t))^L.
\end{equation}
An initial point $\tau_{k_n}^{(0)}$ can be obtained by relaxing constraints (\ref{rHOR1l}) and (\ref{rHOR1n}) and solve the semi-definite relaxation (SDR):
\begin{eqnarray}\label{rHOR6}
\ds\min_{W(t'), R(t'), \tau_{k_n}(t')} F_{[t,\Psi(t)]}\quad
 \mbox{s.t.}\ (\ref{rHOR1d})-(\ref{rHOR1k}), \quad (\ref{rHOR1m}).
\end{eqnarray}
The above iterative procedure terminates at $\ds\frac{1}{g^{(\kappa)}(\tau)}-\frac{1}{\bar{\tau}} < \epsilon$.

\subsection{The second stage: iterative procedure to solve (\ref{rHOR3})}

In optimization (\ref{rHOR3}), it should be noted that at each time $t$, constraint (\ref{rHOR1l}) is involved with $\Psi(t)-t$ nonconvex  matrix rank-one constraints. However, in the dynamic charging problem only the current state and control will be updated. Thus, (\ref{rHOR3}) can be simplified by,
\begin{subequations}\label{rHOR6}
\begin{eqnarray}
\ds\min_{W(t'), R(t')}\ F(P_g(t'))):=\ds \sum_{t'=t}^{\Psi(t)}
\sum_{k\in {\cal G}}f(P_{{g_k}}(t')),\quad \label{rHOR6a}\\
 \mbox{s.t.}\quad (\ref{rHOR1c})-(\ref{rHOR1k}), (\ref{rHOR3c}),\quad t'\in[t,\Psi(t)], \label{rHOR6b}\\
 \mbox{rank}(W(t))=1,\label{rHOR6c}
\end{eqnarray}
\end{subequations}
with only one matrix rank-one constraint (\ref{rHOR6c}).

Following our previous works \cite{Phetal12,STST15,Naetal17,Yeetal17,STA17}, optimization (\ref{rHOR6}) can be solved very efficiently by a nonsmooth optimization algorithm (NOA), which is to proposed to deal with the  matrix rank-one constraint (\ref{rHOR6c}). Optimization (\ref{rHOR6}) is equivalent to
\begin{subequations}\label{rHOR7}
\begin{eqnarray}
\ds\min_{W(t'), R(t')}\ F_{\mu}(W(t'),P_g(t'))):=F(P_g(t'))\nonumber + \\ \mu_2(\Tr(W(t'))-\lambda_{\max}(W(t'))),\quad \label{rHOR7a}\\
 \mbox{s.t.}\quad (\ref{rHOR1c})-(\ref{rHOR1k}), (\ref{rHOR3c}),\quad t'\in[t,\Psi(t)], \label{rHOR7b}
\end{eqnarray}
\end{subequations}
which can be solved by the following iterative procedure:
\begin{eqnarray}\label{rW5}
\ds\min_{W(t'), R(t')} F_{\mu}^{(\kappa)}(W(t'),R(t')):=F(P_g(t'))+\mu_2 (\Tr(W(t'))
\nonumber \\ -(w_{\max}^{(\kappa)}(t'))^H W(t') w_{\max}^{(\kappa)}(t'))\quad
\mbox{s.t.}\quad (\ref{rHOR1c})-(\ref{rHOR1k}), (\ref{rHOR3c}),\quad t'\in[t,\Psi(t)],
\end{eqnarray}
where $w^{(\kappa)}_{\max}(t)$ denotes the normalized eigenvector
corresponding to the eigenvalue $\lambda_{\max}(W^{(\kappa)}(t))$, $\mu_2>0$ is a penalty parameter. The reader is also referred to \cite{Yeetal17} for the convergence proof of the above procedure (\ref{rW5}). The above procedure terminates at
$0\leq \Tr(W^{(\kappa)}(t))-\lambda_{\max}(W^{(\kappa)}(t))
\leq \Tr(W^{(\kappa)}(t))-(w_{\max}^{(\kappa)}(t))^H W^{(\kappa)}(t) w_{\max}^{(\kappa)}(t)
\leq \epsilon$.

In summary, our proposed MPC-based computation for (\ref{PEV1}) is based on two-stage iterative procedure solving (\ref{rHOR2}) for online
coordinating PEV charge $\hat{\tau}_{k_n}(t)$ and solving (\ref{rHOR3}) for online updating the
generated voltage $\hat{V}(t)$ for the generated power $\hat{R}(t)$  by
\begin{equation}\label{Vt}
\hat{V}(t)=\sqrt{\lambda_{\max}(W^{(\kappa)})}w^{(\kappa)}_{\max}(t),
\end{equation}
whenever the solution $\hat{W}(t)$ of SDR (\ref{rHOR3}) is not of rank-one. If $\mbox{rank}(\hat{W}(t))=1$, it is obvious
that $\hat{V}(t)=\sqrt{\lambda_{\max}(\hat{W}(t))}\hat{w}_{\max}(t)$ with the normalized eigenvector $\hat{w}_{\max}(t)$
corresponding to $\lambda_{\max}(\hat{W}(t))$ is the optimal solution of (\ref{HOR1}), which is what we need. Algorithm (NOA) \ref{alg1} is the pseudo-code for the above two-stage dynamic scheduling optimization.

\begin{algorithm}[!t]\caption{Two-stage nonsmooth optimization algorithm for dynamic scheduling (\ref{rHOR1})}\label{alg1}
  \begin{algorithmic}[1]
  \State {\bf Set} $\kappa=0$ solve (\ref{rHOR6}) to obtain the initial point of $\tau_{k_n}^{(0)}$,
  \State {\bf Until}
  $\ds\frac{1}{g^{(\kappa)}(\tau)}-\frac{1}{\bar{\tau}} < \epsilon$, {\bf solve} (\ref{rHOR5}) to find the optimal solution of $(W^{(\kappa+1)}(t'), R^{(\kappa+1)}(t'), \tau_{k_n}^{(\kappa+1)}(t'))$ and {\bf reset} $\kappa+1\rightarrow \kappa$,
  \If{ $\mbox{rank}(\hat{W}(t'))\equiv 1$, $t'\in [t,\Psi(t)]$}
  accept $(W^{(\kappa)}(t'), R^{(\kappa)}(t'), \tau_{k_n}^{(\kappa+1)}(t'))$  as the optimal solution of the nonconvex optimization problem (\ref{rHOR1}),
  \Else{ set $\kappa=0$, $\hat{\tau}_{k_n}(t)=\tau_{k_n}^{(\kappa)}(t')$
   {\bf Until} $ \Tr(W^{(\kappa)}(t))-(w_{\max}^{(\kappa)}(t))^H W^{(\kappa)}(t) w_{\max}^{(\kappa)}(t) \leq \epsilon$,
   {\bf solve} (\ref{rW5}) to find the optimal solution $(W^{(\kappa+1)}(t'), R^{(\kappa+1)}(t'))$ and
  {\bf reset} $\kappa+1\rightarrow \kappa$,}
  \EndIf
  \State {\bf Accept} $(W^{(\kappa)}(t'), R^{(\kappa)}(t'),\tau_{k_n}^{(\kappa)}(t'))$  as the optimal solution of the dynamic scheduling problem (\ref{rHOR1}).
   \end{algorithmic}
\end{algorithm}

\section{Simulation results}
\subsection{Simulation setup}
The SDPs (\ref{rHOR6}), (\ref{rHOR5}) (\ref{rW5}) and (\ref{rHOR7}) are computed using Sedumi\cite{S98} interfaced by CVX \cite{cvx}
on a Core i5-3470 processor. Four power networks from Matpower \cite{ZMT11} are chosen.
The tolerance $\epsilon = 10^{-4}$ is set for the stop criterions.

\begin{figure}[h]
\centering
\includegraphics[width=0.8 \columnwidth]{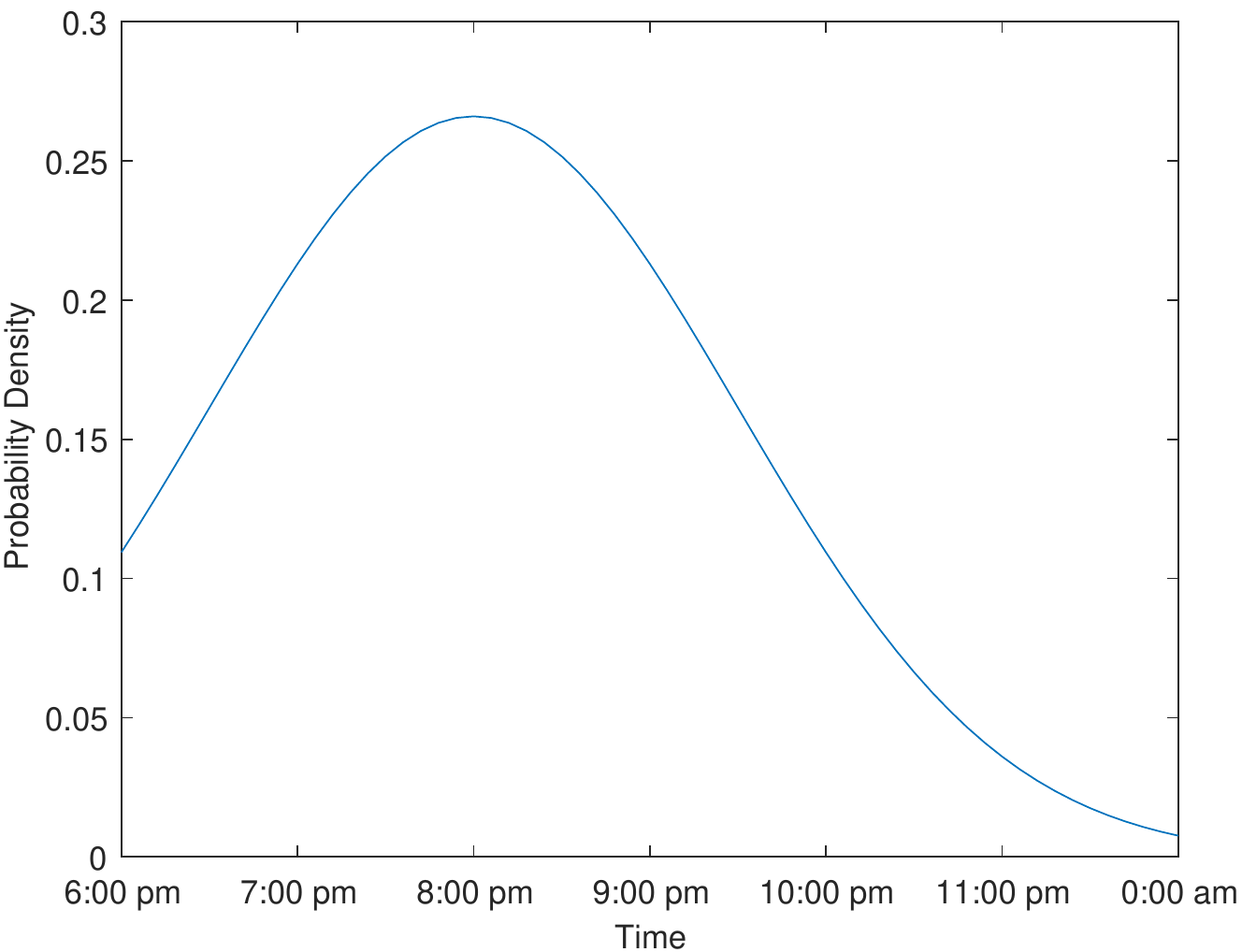}
\caption{The probability density of PEVs' arrivals}
\label{n_arrival}
\end{figure}

Generally, PEVs are charged  after their owners' working hours.
We focus on the charging period from 6:00 pm to 6:00 am of the next day, which is then uniformly
divided into $24$ time slots of $30$ minute length  \cite{JTG13}. Accordingly, the charging time horizon
 is $\clT=\{1, 2, \dots, 24\}$. It is also reasonable to assume that the PEVs arrive during the time period
 from 6:00 pm to midnight. The PEVs must be fully charged after being plugged into the grid. The arrival times of
 PEVs are assumed to be independent and are generated by a truncated  normal distribution $(20,1.5^2)$, which is depicted by Fig. \ref{n_arrival}.

We assume that the PEVs are Tesla Model S's, which have a battery capacity of  100 KWh \cite{Tesla}.
The SOC of all PEVs is set as 20\%. The structure and physical limits of the considered grids are given in the Matpower
library \cite{ZMT11} together with the specific cost functions $f(P_{{g_k}}(t))$.

 Without loss of generality,  PEV loads are connected at the generator buses, which means each generator bus will serve as a charging station.

\begin{figure}[h]
\centering
\includegraphics[width=0.8 \columnwidth]{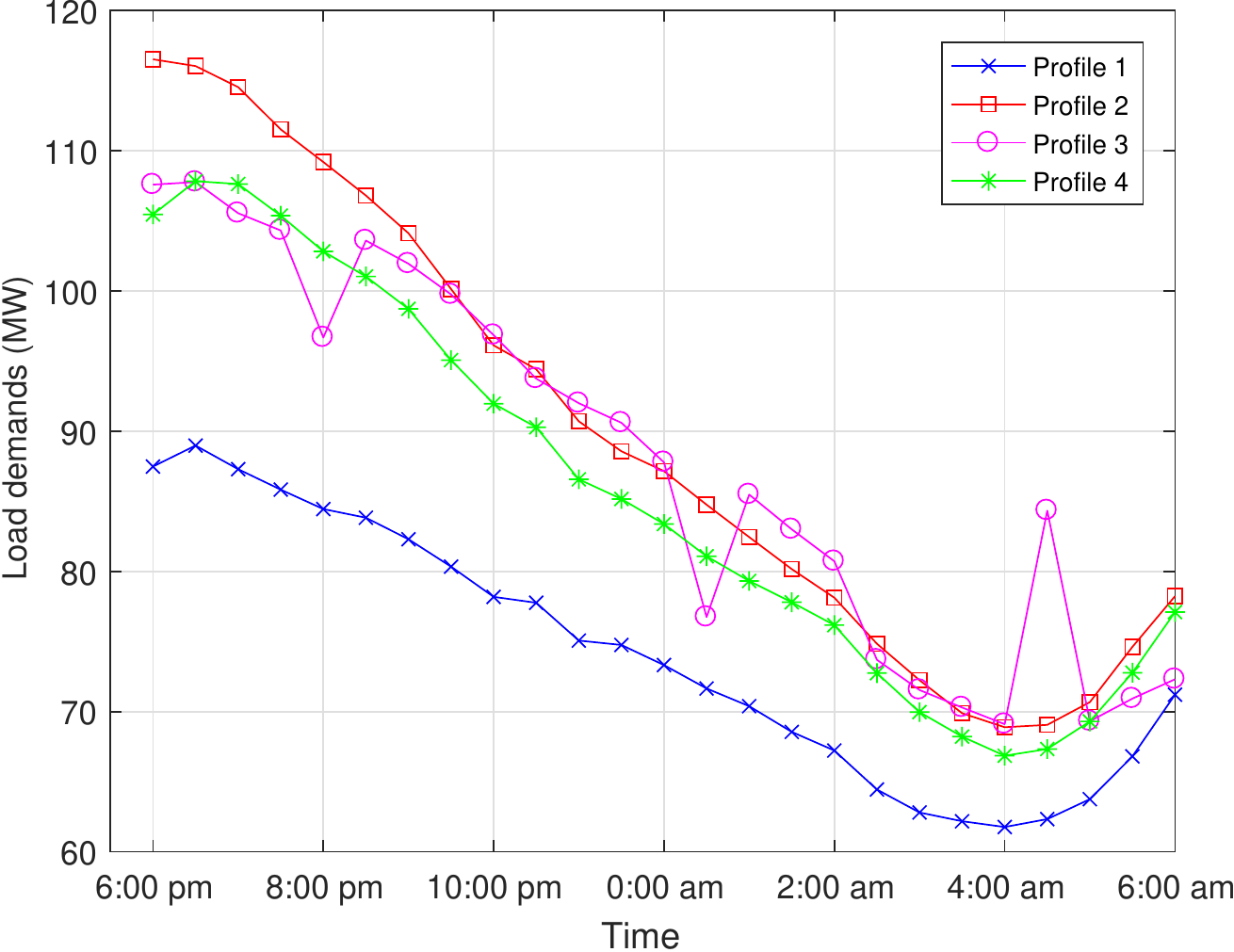}
\caption{Residential load demands of four profiles}
\label{Load_demand}
\end{figure}

\begin{figure}[h]
\centering
\includegraphics[width=0.8 \columnwidth]{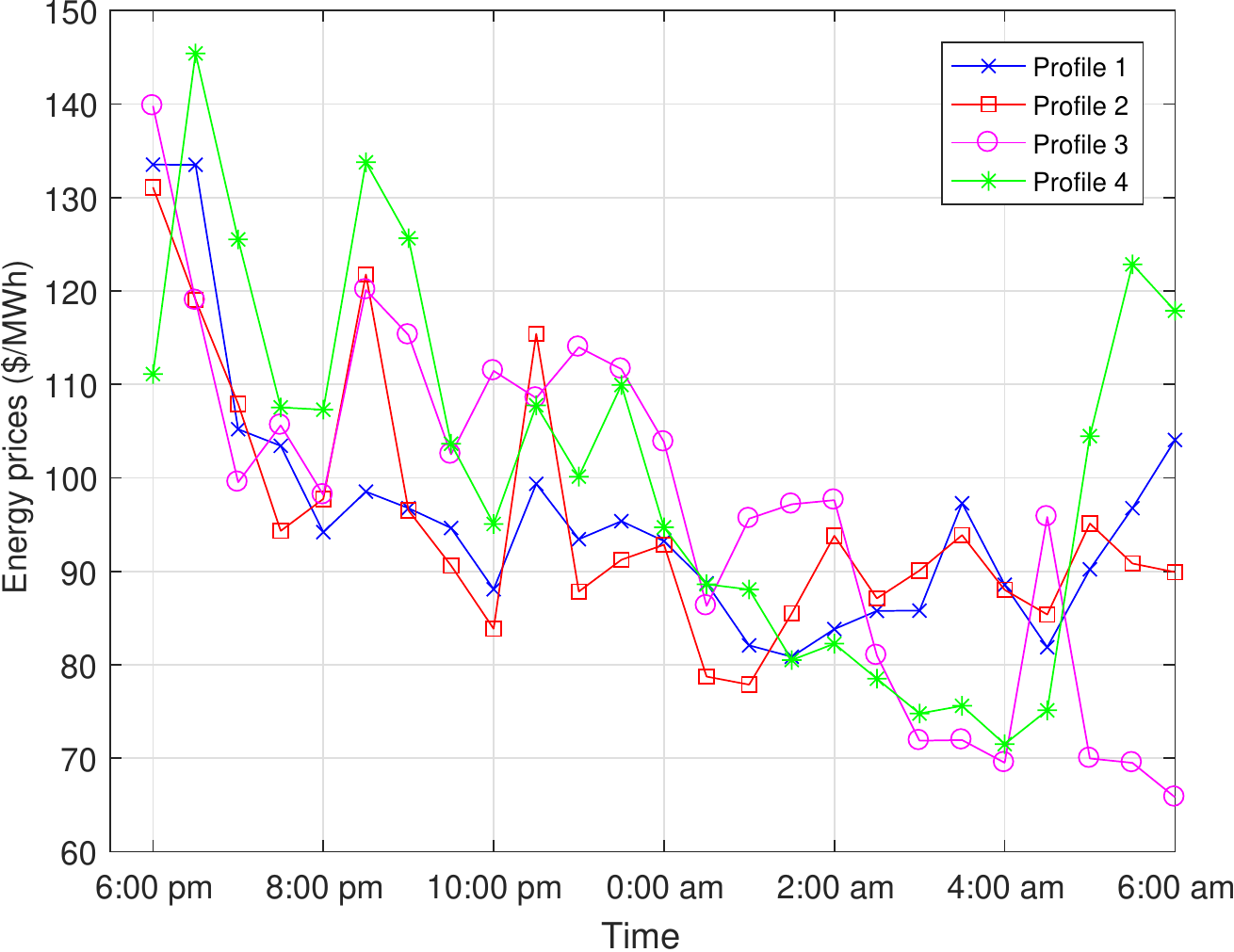}
\caption{Energy prices for four profiles}
\label{Energy_price}
\end{figure}

The price-inelastic load $P_{l_k}(t)$ is calculated as
\begin{equation}
P_{l_k}(t) = \frac{l(t)\times \bar{P}_{l_k}\times T}{\sum_{t=1}^{24} l(t)}, \quad t \in \clT,
\end{equation}
where $\bar{P}_{l_k}$ is the load demand specified by \cite{ZMT11} and $l(t)$ is the residential load demand
taken from \cite{NSW_data}. Four profiles are taken from different days in 2017. Profile 1 is the residential load and energy price from 6:00 pm on 7th May to 6:00 am on 8th May, Profile 2 is from 6:00 pm on 7th June to 6:00 am on 8th June, Profile 3 is from 6:00 pm on 7th July to 6:00 am on 8th July, and Profile 4 is from 6:00 pm on 7th August to 6:00 am on 8th August. Fig. \ref{Load_demand} and Fig. \ref{Energy_price} provide the residential load demand and energy price for these profiles.

\subsection{MPC-based dynamic scheduling computational results}

\subsubsection{Four network simulation}
We test MPC-based online computation for  Case9, Case14, Case30 and Case57 from \cite{ZMT11} and profile 2
of the residential data. The information on these networks is given  in Table. \ref{Case_inform}, where the first column is the name of network,
the second column indicates the numbers of buses, generators and branches.
The dimension of $W(t)$ is given in the third column, while the total number of PEVs is shown in the last column.
\begin{table}[h]
    \centering
    \caption{Information on four networks}
    \begin{tabular}{ccccc}
    \hline
    & Buses/Generators/Branches & Dim. of $W(t)$ & PEVs\\
    \hline
    Case9 & 9/3/9  & $\mathbb{C}^{9\times 9}$ & 126 \\
    Case14& 14/5/20 & $\mathbb{C}^{14\times 14}$ & 210  \\
    Case30& 30/6/24 & $\mathbb{C}^{30\times 30}$ & 252  \\
    Case57& 57/7/80 & $\mathbb{C}^{57\times 57}$ & 294  \\
    \hline
    \end{tabular}
\label{Case_inform}
\end{table}
The computational results are summarized in Table \ref{mpc_network}.
\begin{table}[h]
    \centering
    \caption{MPC-based two-stage results}
    \begin{tabular}{cccccccc}
    \hline
     &Binary variables &$\mu_1$ & $\mu_2$ &Stage-1 &Stage-2& Time(s) \\
    \hline
    Case9& 1512 &1 &10 & 23854.7&23858.3&  25.5 \\
    Case14& 2520 &1 &- & 53431.1&53431.1& 18.5 \\
    Case30& 3012 &1 &10 & 5634.9&5639.6&  35.5 \\
    Case57& 3528 &10 &10 & 87490.7& 87502.7 & 89.7\\
    \hline
    \end{tabular}
\label{mpc_network}
\end{table}
Again, the first column is the network name. The second column presents the number of binary variables $\tau_{k_n}(t')$ in (\ref{HOR1}). The value of
the penalty parameter $\mu_1$ in (\ref{rHOR5}) and $\mu_2$ in (\ref{rW5}) are given in the third column and forth column, respectively. The computational value of the first stage and the second stage are respectively provided in the fifth column and sixth column. In the last column, the average running time is presented.


\subsubsection{Four residential profile simulation}
 We consider Case30 together with four different residential profiles.
 The computational results are provided in Table \ref{mpc_load}, whose format is similar to Table \ref{mpc_network}.
\begin{table}[h]
    \centering
    \caption{MPC results for Case30 with four different residential profiles}
    \begin{tabular}{cccccccc}
    \hline
     &Binary variables &$\mu_1$ & $\mu_2$ &Stage-1 &Stage-2& Time(s) \\
    \hline
    Profile 1& 3012 &1 &10 & 7834.5&7836.7&  34.8 \\
    Profile 2& 3012 &1 &10 & 5634.9&5639.6& 35.5 \\
    Profile 3& 3012 &1 &10 & 8632.5&8636.4&  36.7 \\
    Profile 4& 3012 &1 &10 & 6541.3& 6545.1 & 35.2\\
    \hline
    \end{tabular}
\label{mpc_load}
\end{table}
\section{Conclusions}

Joint PEV charging scheduling and power control for  power grids
to serve both PEVs at a competitive cost  and residential power demands at a competitive operating cost
is very difficult due to the random nature of PEVs' arrivals and  demands. We have proposed a novel
and easily-implemented  MPC-based two-stage computational algorithm that can achieve a globally optimal solution.

\bibliographystyle{ieeetr}
\bibliography{EV_BIB}
\end{document}